\documentclass[11pt]{article}
\usepackage{jheppub}
\usepackage{amsmath}
\usepackage{amssymb}
\usepackage{braket}
\usepackage{bbold}
\usepackage{pifont}
\usepackage{color}
\usepackage{dsfont}
\usepackage{hyperref}
\usepackage{subcaption}

\newcommand{\beqn}{\begin{eqnarray}}
\newcommand{\eeqn}{\end{eqnarray}}

\begin{document}

\title{A Novel Approach To Particle Representations}
\author{Brage Gording}
\abstract{
	
	This paper proposes a new approach to  deriving a finite particle content, suitable for the construction of a gauge theory. Specifically, the outlined construction generates a finite set of irreducible gauge representations, which are interpreted as describing a full set of elementary particles. These representations are constructed from endofunctions between restricted representations of some symmetry group $G$ acting on some space $V$. As a proof of concept, we show how a set of irreducible representations arise as endofunctions on the vector space $V=\mathds{C}^8$ equipped with the exceptional Lie group $G=G_2$ as its symmetry group. We discuss how the irreducible representations of our simple example compare to the various particle types of the Standard Model. The process through which the particle content is constructed yields adjoint, fundamental, and Higgs-like representations, thereby reproducing the essential types of particle transformations seen in the Standard Model. In particular we focus on the discrimination of gauge structures and the natural appearance of Higgs-like representations. Avenues to generalizing the construction are considered, and some inevitable consequences are discussed. We conclude by comparing our results to those of non-commutative geometry, commenting on key similarities and differences between the two approaches.}


\begin{flushright}
	\hfill{MPP-2020-63} \vspace{20mm}
\end{flushright}

\maketitle
\flushbottom



\section{Introduction}
\label{introduction}

The Standard Model of particle physics is a gauge theory describing subatomic interactions to incredible accuracy, Ref.~\cite{Sirunyan:2020ztc,Azzi:2019yne}. Specifically, its action is constructed from fields, describing particle excitations, that transform under irreducible representations of both the global Lorentz group, SL(2,$\mathds{C}$),%
\footnote{The Lorentz group is actually SO(1,3), but fermions lie in representations of SL(2,$\mathds{C}$) which is the double cover of SO(1,3).}
and the local Standard Model gauge group $G_{\text{SM}}=\text{SU}(3)\times\text{SU}(2)_L\times\text{U}(1)_Y$. Despite the theory's predictive power, it does not explain the origin of the particular choice of representations of $G_{\text{SM}}$ which are realized in the Standard Model. In other words, the model offers no explanation for its particle content. In particular, the model contains representations which appear in multiplicity, i.e. the generations of fermionic particles, while other representations are not realized at all, i.e. the absence of right-handed SU(2) doublets. 

Many attempts have been made at explaining the Standard Model gauge representations, notably Grand Unified Theories (GUT's) attempt to explain representations of particles from the choice of gauge groups themselves. GUT's are motivated by an approximate gauge coupling unification which appears if the running of gauge couplings are extrapolated to very high energies, Ref.~\cite{Georgi:1974sy,Weinberg:1974yx,Georgi:1974yf}. The simplest such model is also the original GUT, proposed by Georgi and Glashow in 1974. This model is based on SU(5), which contains $G_{\text{SM}}$ as a subgroup. The standard approach of GUT's is then to 
endow all additional gauge fields, corresponding to generators not in $G_{\text{SM}}$, with large masses through spontaneous symmetry breaking. Thus the effects of the additional gauge bosons only become relevant at energies approaching these masses, and we recover the Standard Model as the low energy theory. This simplest SU(5) GUT has, however, already been ruled out by experiments on proton decay, Ref.~\cite{Miura:2016krn}.

The Pati-Salam model attempts to explain the SU(2)$_L$ stucture of the Standard Model in a similar approach by re-introducing left-right chiral symmetry. This is accomplished via the introduction of an additional gauge group, SU(2)$_R$, which only acts on right handed particles. Unlike SU(5), the gauge-mediated proton decay is avoided in the Pati-Salam model based on the gauge group SU(4) $\times$ SU(2)$_R$ $\times$ SU(2)$_L$, Ref.~\cite{Pati:1973uk,Pati:1974yy,Mohapatra:1974hk}. Then $G_{\text{SM}}$ is obtained spontaneously breaking the SU(4)$\times$SU(2)$_R$ group down to the SU(3)$\times$U(1)/$\mathbf{Z}_3$ subgroup of the Standard Model. This model is not a GUT, as it contains multiple distinct group structures, but can be embedded into the GUT based on SO(10).%
\footnote{Note that while this GUT is referred to as ``SO(10)" by physicists, the Lie group used is actually Spin(10), the double cover of SO(10).}

The SO(10) GUT has some nice features, such as the 16-dimensional spinor representation that exactly incorporates one generation of the Standard Model fermions (including the unobserved right-handed neutrino), Ref.~\cite{Georgi:1974my}. However, spontaneously breaking the gauge group down to $G_\mathrm{SM}$ requires a Higgs sector with representations of large dimensions. As a consequence, already the minimal model contains more than 100 new fields. Further, GUT's only explain the appearance (or absence) of representations, not their multiplicity, and thus do not explain the appearance of the three generations of fermions. For more information on GUT's, we refer the interested reader to Ref.~\cite{Georgi:1981vf,Grinstein:1982um,Masiero:1982fe}.

\subsection{Motivation}
\label{motivation}

These shortcomings of minimal GUT's motivates us to consider a change in perspective. This is based primarily on two reasons: first we wish to find a construction that yields a particle content, not just a set of representations; and second we seek a way to generate the Standard Model particle representations without requiring the addition of a multitude of new particles. As such, we seek a construction method which can yield all necessary representations of the Standard Model. That is, we require the generation of adjoint, fundamental, and singlet representations. Note that for the purpose of this paper we focus only on the gauge group structure, not the Lorentz structure of the Standard Model. This is also the standard approach of GUT's. We will discuss Lorentz structures briefly in section \ref{comments}. Further, the purpose of this paper will not be to reproduce the Standard Model particle content, only to introduce a construction method that generates the types of particle representations seen in the Standard Model.

The essence of the construction is to consider some space $V$ together with a symmetry group $G$, and then look at maps between irreducible representation spaces of subgroups of $G$. This is somewhat similar to the construction of GUT's, as we are employing a larger group $G$ which contains various subgroups of the Standard Model gauge group. However, we will neither require $G_{\text{SM}}\subset G$ nor any spontaneous symmetry breaking to recover the $G_{\text{SM}}$ group structure. The idea is based on the following observation:
\begin{enumerate}
	\item[]
	Consider a vector space $X_1$ transforming as $X_1\to a_1^{-1}X_1$ where $a_1\in\rho_1(H_1)$, for some group $H_1$ and representation $\rho_1:H_1\to GL(X_1)$. Similarly consider a vector space $X_2$ transforming as $X_2\to a_2 X_2$ where $a_2\in\rho_2(H_2)$, for some group $H_2$ and representation $\rho_2:H_2\to GL(X_2)$. Then the set of maps $M:X_1\to X_2$ transform as $M\to a_2 M a_1$, that is in the fundamental representation of both $H_1$ and $H_2$. Similarly, endofunctions on $X_i$ transform in the adjoint representation of $H_i$.%
\end{enumerate}

Now, all Standard Model fields transform either in the fundamental or adjoint representations of the gauge group $G_{\text{SM}}$. Therefore, all gauge transformations of the Standard Model are realizable as transformations of maps between spaces endowed with different symmetry structures.%
\footnote{Indeed in Ref.~\cite{Gording:2019srz} it was shown that the entire Standard Model could be embedded in M(32,$\mathds{C}$), the 1024 $\mathds{C}$-dimensional space of 32$\times$32 complex matrices, with only a 16 $\mathds{C}$-dimensional subspace lying outside the Standard Model particle content. However in this construction there was no fundamental reason for the appearance of Standard Model gauge groups, and no derivation of particle content.}
Consequently we wonder whether there may be an underlying structure from which the Standard Model particle content may be realized as a set of endofunctions. To investigate whether such a construction would even be possible we must first answer a more general question:
\begin{enumerate}
	\item[]
	Given a space $V$, the symmetry group $G:=\text{Sym}(V)$, and the space of endofunctions M($V):V\to V$; can we find a direct sum decomposition of M($V$) into subspaces $S_i$, such that each $S_i$ transforms under irreducible restricted representations of $G$?
\end{enumerate}
We consider a direct sum decomposition of the matrix algebra M($V$) as we may then span the space M($V$) by irreducible representations; this follows immediately from M($V$) being itself a vector space. Therefore such a decomposition of M($V$) would make it to understand transitions between elements in $V$ as a set of particle representations. Throughout this paper linear independence of elements of the matrix space M($V$) will therefore be central to the discussion.

Note that we do not demand that this direct sum decomposition be unique. Uniqueness is an attractive feature of a construction, and further development of this construction may indeed yield a process which generates a unique particle content. However, at the current stage, the purpose of this paper is only to show that the above construction can indeed yield a complete set of particle representations: including everything from adjoint representations, i.e. gauge fields, to representations with properties unique to the Higgs sector of the Standard Model. Further, we must also acknowledge that uniqueness may be too restrictive an assumption on the construction. Indeed there are situations in which fermionic degrees of freedom can be expressed in terms of bosonic ones, or visa-versa, Ref.~\cite{vonDelft:1998pk,Senechal:1999us}. Therefore, at the present moment, we will only demand that our construction yields a finite set of consistent decompositions of M($V$), this is trivially the case for any finite group $G$.

In this paper we show that for the specific pair $(V,G)=(\mathds{C}^8,G_2)$, the above described decomposition of M($V$) exists under a very minimal set of assumptions, detailed in section \ref{axiomatisation}.

\subsection{Organization of Paper}

In section \ref{smreps} we emphasize some properties of the irreducible representations of the Standard Model, for easier comparison with our own results. Next, in section \ref{axiomatisation}, we outline a set of construction principles chosen to ensure the appearance of representations necessary to construct a gauge theory of conserved charges. These principles are very minimal and detailed in the subsections \ref{particles as maps} - \ref{paprelation}. In section \ref{derivation} we show how, for the particular choice of ($\mathds{C}^8,G_2$), we can express the maps between SU(3) and SU(2) decompositions of $\mathds{C}^8$ as spanned by a set of irreducible representations. We identify these representations with a particle content.

The particle content itself is analyzed, and compared to representations of Standard Model particles in section \ref{comments}. Specifically we show how we obtain three distinct types of representations, and compare them to the representations of Standard Model particles of type: gauge boson, fermion, and Higgs doublet. In section \ref{vectoralgpurespin} we discuss the relationship between vector spaces and algebras in our construction. We then discuss possible generalizations of the construction in section \ref{generalization}. Even with these possible generalizations we find that there are distinct features which are inevitable consequences of our approach. In section \ref{predictions}, we show how these features would imply deviations if attempting to derive the Standard Model particle content via our proposed construction, and therefore yield testable predictions. In section \ref{noncommutativegeometry} we compare our results to those found in non-commutative geometry, and discuss the appearance of concepts that are central to both constructions. The paper is summarized in section \ref{conclusion}.

\section{Representations in the Standard Model}
\label{smreps}
We here review a few properties of the particle representations in the Standard Model. Our construction applied to the simple pair ($\mathds{C}^8$,$G_2$) will be presented in section \ref{derivation}. The comparison between these two sets of representations is presented in section \ref{comments}.

\subsection{Spinor Representations}
\label{fundreps}

Consider the fundamental representations of the group SL(2,$\mathds{C}$)$\times$SU(3)$\times$SU(2) present in the Standard Model, ignoring U(1) charges. We observe Lorentz spinors in the distinct fundamental representations
\beqn
\label{smspinorreps}
(2,3,2),(2^*,3^*,2^*),(2,3^*,1),(2^*,3,1),(2,1,2),(2^*,1^*,2^*),(2,1,1),(2^*,1^*,1^*);
\eeqn
where (a,b,c) denotes a particle in the $a$ representation of SL(2,$\mathds{C}$), $b$ representation of SU(3), and $c$ representation of SU(2), and $^*$ denotes the complex conjugate representation, also known as the anti-fundamental representation, Ref.~\cite{Burgess:2007zi}.%
\footnote{It is conventional to denote these conjugate representations of SU(N) as $\bar{N}$ instead of $N^*$. However, as we will see, in this construction the notion of complex conjugation denotes both basis elements and representations, and therefore is kept explicitly.}
So a $2$ of SL(2,$\mathds{C}$) is a left-handed spinor while $2^*$ is a right-handed spinor. For example, the $(2,3,2)$ representations describes a left-handed SU(3) triplet which is also a SU(2) doublet, i.e. a left-handed up-down quark pair. Similarly the $(2^*,3,1)$ representation describes a right-handed SU(3) triplet which is a singlet of SU(2), i.e. an up- or down-type quark.

Combining the irreducible singlet and triplet representations of SU(3), we obtain the reducible representations
\beqn
\label{lhp}
(2,\,1\oplus3,\,2)\quad\text{ and complex conjugate }\quad(2^*,\,1^*\oplus3^*,\,2^*)
\eeqn
as well as
\beqn
\label{rhp}
(2,\,1^*\oplus3^*,\,1)\quad\text{ and complex conjugate }\quad(2^*,\,1\oplus3,\,1^*).
\eeqn
Note that singlet states do not transform. This trivially implies that, as representations, $1^*\sim1$; however, we will see that keeping the $^*$ notation explicit will be useful for later construction. Thus, only left handed particles (and by complex conjugation right handed antiparticles) transform in the fundamental representation under SU(2). This property is usually referred to as the theory being chiral, due to the different gauge structures of left and right handed spinors.

In our construction we will see this difference in gauge representations as a property of the pair ($V,G$). Thus in our construction the property of a theory being ``Chiral" is then a specific case of a more general phenomenon. We will refer to the general phenomenon instead as a non-uniform realization of gauge representations, and we will show the appearance of this phenomenon in section \ref{derivation}. Note that we will use the terminology of the Standard Model being ``Chiral" to describe the appearance of only left handed SU(2) doublets in the Standard Model.

\subsection{The Higgs Doublet}
\label{smhiggsrep}

Although we observe particles in all the representations listed in (\ref{smspinorreps}), the Standard Model itself does not explicitly contain all these representations in the action. Indeed only half the representations in (\ref{smspinorreps}), along with their duals, appear explicitly.%
\footnote{Dual representations of unitary groups naturally implies the hermitian conjugate representation.}
That is, for a spinor we need only one of the representations $R$ or $R^*$, along with its dual, to construct all relevant terms in the action. This is not the case for the Standard Model Higgs doublet, for which we require its representation $R$, its conjugate representation $R^*$, and the dual representations $R^\dagger$ and $\left(R^{*}\right)^\dagger$ to construct all the necessary Yukawa interactions.

To construct the appropriate Yukawa interactions of the Higgs doublet we require the explicit appearance of both the $(1,1,2)$ representation and its complex conjugate, $(1^*,1^*,2^*)$. These representations are used to form invariants between left and right handed fermions, i.e. particles of different gauge representations, Ref.~\cite{Buchmuller:2006zu}. Additionally, the Higgs doublet is further unique in the Standard Model as the only field which transforms in the fundamental representation of its gauge groups while at the same time being its own antiparticle. In section \ref{comments} we will see that irreducible representations with these crucial properties arise naturally from our proposed construction.

\section{Principles of Construction}
\label{axiomatisation}

The purpose of this paper is to show how a set of irreducible gauge representations appear as the constituent subspaces of a direct sum decomposition of the space of maps on some pair $(V,G)$. Of course there are an infinite number of ways to decompose said space of maps; however, not all such decompositions will admit a particle interpretation. Therefore, we need to ensure that we obtain a decomposition of the space of maps M($V$) where the subspaces transform under irreducible representations of the gauge group. This requires the imposition of construction principles that restrict the specific form of the subspaces, i.e. criteria for the decomposition of M($V$) into subspaces. 

The need for these construction principles is obvious: an algebra is only an abstract mathematical concept. Therefore the construction principles are a guide of how to represent the abstract algebra M($V$) in terms of physical objects, i.e. particles. We present here a set of construction principles, formulated in terms of three ``conditions" on the algebra, which ensure the existence of a relevant direct sum decomposition of the space of maps on $(\mathds{C}^8,G_2)$.%
\footnote{Note that if one wishes one may separate, for example, condition (1) into a set of multiple construction principles. In other words, there is nothing about specifically three conditions that is fundamental to this construction. Rather, it is the total set of construction principles which is essential, not the way in which they are grouped. Here, we choose to group the all construction principles into three conditions for simplicity and brevity.}

We intentionally employ only very mild assumptions and general construction principles. Indeed, we will show in the following subsections that the construction principles detailed below are nothing more than the most basic requirements for a gauge theory of elementary particles. Having such general construction principles will prevent us from ensuring uniqueness of the decomposition of M($V$). While the lack of uniqueness is not an appealing trait for a general construction, the purpose of this paper is not to present the construction in its full generality; the purpose instead being a proof of concept of a much more general construction idea.

Therefore, we choose to only employ the mildest assumptions in our construction principles. This ensures we do not impose any restrictions for the general construction which are only relevant to our particular choice of $(\mathds{C}^8,G_2)$. We comment on the choice of principles in section \ref{VGchoice}.

\begin{enumerate}
	\item Particle representations are maps between decompositions of $V$, forming sets that transform irreducibly under restricted representations of non-trivial, continuous, subgroups of $G$. Every such group has an associated Lie algebra which describes a set of particles transforming in the adjoint representation.\label{cond1}

	\item Distinguishable particles appear as linearly independent elements of M($V$).\label{cond2}
	
	\item The decomposition of M($V$) into a set of irreducible representation spaces must be invariant under complex conjugation.\label{cond3}
\end{enumerate}

\subsection{Particles as Maps}
\label{particles as maps}

In the Standard Model, particles transform under irreducible representations of $G_{\text{SM}}$, and every gauge group comes with an associated gauge field. This is not a unique property of the Standard Model. Rather, this is a core construction principle of any gauge theory of elementary particles, as reducible representations can always be constructed from irreducible ones. Thus condition \ref{cond1} is equivalent to the statement that we are looking for a gauge theory of elementary particles.

The particular choice of the exceptional Lie group $G_2$ is primarily because this group contains as independent subgroups $SU(3)$ and $SU(2)$. In fact SU(3) and SU(2) are the only non-abelian subgroups of $G_2$, with their Lie algebras appearing as independent subalgebras of the Lie algebra $\mathfrak{g}_2$ of $G_2$,
\beqn
\mathfrak{su}(3)\oplus \mathfrak{su}(2)\subset \mathfrak{g}_2\quad\text{ and }\quad \mathfrak{su}(3)\cap\mathfrak{su}(2)=\{0\}.
\eeqn
Working with similar gauge groups to those in the Standard Model will ease comparisons in section \ref{comments}. We stress again that there is no attempt in this paper to reproduce the Standard Model particle content, only to propose a new construction principle capable of generating the necessary representations of gauge theories.

\subsection{Linear Independence}

In our construction the notion of linear independence of basis elements will play a crucial role for identifying particle states. However, this is not a construction principle unique to our setup. Indeed, linear independence is crucial for particle identification in any gauge theory, i.e. like the Standard Model. To make this point explicit, consider a triplet of red, green, and blue quarks. For ease of calculation one often allocates specific basis elements in $\mathds{C}^3$ to each colour, Ref.~\cite{GellMann:1962xb}. However, such an allocation is arbitrary. In general we can at most say that the set of quark ``colours" span the entire 3-dimensional fundamental representation of SU(3), i.e. that there exists three linearly independent charges/particles which are related by SU(3) transformations. This statement may in general be phrased as: elementary particles are such that they may not be expressed as a linear combination of other elementary particles.%
\footnote{Note that we treat quarks of different colour as distinct elementary particles. This is in direct agreement with how a left handed electron and the corresponding electron-neutrino are treated as distinct elementary particles, even though they are simply two distinct SU(2) charges of a left handed lepton doublet.}
Of course when particles lie in distinct spaces, i.e. like quarks which lie in $\mathds{C}^3$ and gluons which lie in M($\mathds{C}^3$), this point is moot. However, in the construction that will follow all particles, both in the adjoint and fundamental representations, lie in the space M($V$). Consequently, in this construction a complete set of elementary particles naturally become interpreted as a basis of the space in which they exist.

Note that for consistency this implies that each generation of particles must appear as linearly independent elements of M($V$). Specifically this would mean we would need three linearly independent elements of M($V$) all with the same gauge representations to describe the electron, muon, and tau leptons. Similarly we would need 9 linearly independent elements to describe the gauge structures of up, charm, and top quarks. In this way the construction can explain the multiplicity of representations (i.e. generations of particles) as sets of linearly independent subspaces in the same representations of the gauge groups.

Of course, this notion of linear independence must be combined with our construction of particle representations realized as maps between irreducible restricted representations, section \ref{particles as maps}. Further, for definiteness we will find a direct sum decomposition for the whole space in terms of irreducible representation spaces.%
\footnote{This point is crucial, because it demands of fixed particle content for any choice of decomposition of $V$. Indeed, for a large enough space $V$ and group $G$ one could find almost any set of linearly independent representation spaces of M($V$).}
Thus we must find a decomposition of M($V$) into a direct sum of subspaces $S_i$ where each subspace transforms in an irreducible representation of SU(3) and SU(2) as subgroups of $G_2$.

\subsection{Particle-Antiparticle Relation}
\label{paprelation}

Finally, the only other condition we will impose on the decomposition of our space of maps is that the decomposition remains invariant under complex conjugation. Specifically this means that if we find some decomposition into irreducible representation spaces $\{S_i\}_{i\in I}$, i.e.
\beqn
\label{mdecomp}
\text{M}(V)=\bigoplus_{i\in I} S_i,
\eeqn
then for every $S_i$ there must exists some unique $S_j$ such that $S_i^*=S_j$.

For spaces where $S_i^*=S_i$ we have particles, i.e. such as gauge bosons, which are their own antiparticles; if $S_i^*\neq S_j$ we have fields with observationally distinct antiparticles, i.e. like quarks and anti-quarks. In other words, invariance of the decomposition (\ref{mdecomp}) under complex conjugation is just the statement that every particle has a corresponding antiparticle.%
\footnote{Note that in the Standard Model we often interpret the dual, i.e. hermitian conjugate, representations to denote anti-particle states. This arbitrary, as for finite-dimensional representations of a compact Lie group the dual and conjugate representations are isomorphic, Ref.~\cite{CompactLieGroups}.}

\section{The Particle Content from $(\mathds{C}^8,G_2)$}
\label{derivation}
\begin{figure}[h]
	\centering
	\includegraphics[scale=0.4]{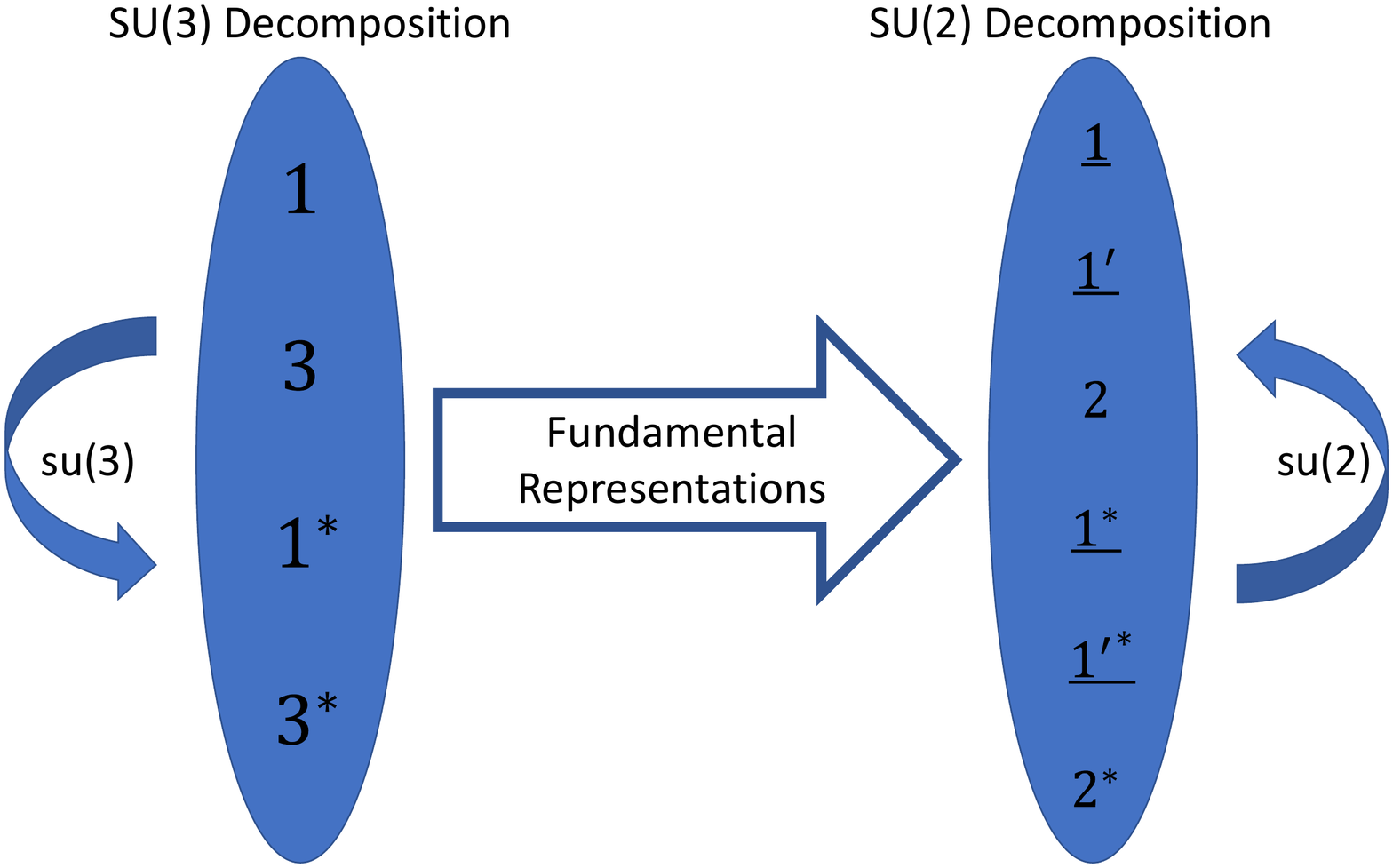}
	\caption{Pictographic illustration of decomposition}
	\label{fig: decomposition}
\end{figure}

As already mentioned in section \ref{particles as maps}, we will be focusing on the decompositions of $V$ under the only two non-abelian subgroups SU(3) and SU(2) of $G_2$. The following subsections involve technical, but straight forward, calculations which illustrate the power of the simple construction principles of section \ref{axiomatisation}. Pictographically the decomposition can be illustrated as shown in \textbf{Figure \ref{fig: decomposition}}.

Statements about linear independence follow immediately from the analysis in Appendix \ref{octonion appendix}. The results of the decomposition are presented in section \ref{comments}.

\subsection{The Adjoint Representations}
\label{nonabrep}

The SU(3) subgroup of $G_2$ splits our 8-dimensional complex vector space into 4 irreducible representation spaces:
\beqn
\label{su3structure}
V = \big(1\oplus 3\big)\oplus \big(1^*\oplus3^*\big),
\eeqn
where the complex conjugate denotes both the subspaces' transformations under SU(3), and that the different subspaces are themselves related by complex conjugation. Here $1$ is a singlet and $3$ is a triplet of SU(3), see Appendix \ref{octonion appendix}. Similarly we have the irreducible representation spaces of SU(2)
\beqn
\label{su2structure}
V = \big(\underline{1}\oplus \underline{1}'\oplus 2\big)\oplus \big(\underline{1}^*\oplus {\underline{1}'}^*\oplus2^*\big).
\eeqn
where we have introduced both an underline and a prime notation on our SU(2) singlet states to emphasize that, in general, they do not correspond to the singlet states of SU(3), Appendix \ref{octonion appendix}. Note that the only difference between the singlet states is that they constitute different subspaces of $V$, this will be important to ensure linear independence of our maps, as specified by condition \ref{cond2}. For brevity we will group the singlet states of SU(2) into a four dimensional complex vector space $1_2$, the subspace $2$ is the doublet of SU(2).

Considering maps from the vector space $V$ to itself it is clear that the Lie algebras corresponding to the SU(3) and SU(2) Lie groups are themselves realized as endofunction on $V$ that preserve the irreducible subspaces. I.e.
\beqn
\mathfrak{su}(3):\big(1\oplus 3\big)\oplus \big(1^*\oplus3^*\big)\to\big(1\oplus 3\big)\oplus \big(1^*\oplus3^*\big),
\eeqn
and similarly
\beqn
\mathfrak{su}(2):1_2\oplus2\oplus2^*\to1_2\oplus2\oplus2^*
\eeqn
where the Lie algebras satisfy their respective commutation relations, and transform in the adjoint representation of their associated Lie groups.

Further, as $\mathfrak{su}$(3) and $\mathfrak{su}$(2) are linearly independent subalgebras of $\mathfrak{g}_2$ they are linearly independent subspaces of M($V$). This is because the representation of $\mathfrak{g}_2$ acting on $V$ is naturally a subset of GL($V)\subset\text{M(}V$). Thus we derive our gauge boson representations as the maps that preserve the irreducible representations of our SU(N). Clearly as any generator $\tau\in\mathfrak{su}$(3), $\mathfrak{su}$(2)$\subset \mathfrak{g}_2$ satisfies $\tau^*=-\tau$, Ref.~\cite{Furey:2018yyy},
we find that gauge bosons are their own antiparticles, as expected.%
\footnote{This uses the physics convention of the fundamental representation of SU(N) generated by hermitian elements. In the mathematics convention where SU(N) is generated by skew-hermitian elements we have instead $\tau^*=\tau$.}
Note, we have only considered the groups SU(N), even though the decompositions (\ref{su3structure}) and (\ref{su2structure}) are left invariant under the respective U(N) groups. We will comment on the appearance of U(1) symmetries in section \ref{comments}.

\subsection{Fundamental and Singlet Representations}
\label{fundandsingreps}

Now for any $\mathfrak{su}$(2) and $\mathfrak{su}$(3) we choose, such that $\mathfrak{su}$(2)$\nsubseteq\mathfrak{su}$(3), we have that the subspaces $3\subset V$ and $2\subset V$, or equivalently $2^*\subset V$, are linearly independent, see Appendix \ref{octonion appendix}. So what maps from $3$ to some $W\subset V$ are such that combined with their complex conjugate maps they are independent from the Lie algebras $\mathfrak{su}$(2) and $\mathfrak{su}$(3)?

Let $\widetilde{W}\subset V$ be such that $3\,\cap\, W=\widetilde{W}$. Then if $\widetilde{W}\neq\{0\}$ there exists some $\lambda\in\mathds{C}\otimes\mathfrak{su}$(3) such that $\lambda:3\to\widetilde{W}$ and $\lambda:3^*\to\widetilde{W}^*$. It is then clear that $\mathfrak{su}$(3) is not linearly independent from the set of maps $3\to \widetilde{W}\subset W$ and $3^*\to \widetilde{W}^*\subset W^*$. Thus the space of maps that transform in the fundamental representation of SU(3) and are  linearly independent from $\mathfrak{su}$(3) is the subspace
\beqn
F_3:3\to J\subset V, \text{ for some $J$ such that } 3\,\cap\, J=\{0\},
\eeqn
along with its complex conjugate space. Consider the singlet space $1_2$ of SU(2). This space is linearly independent form our SU(3) triplet spaces, $1_2\cap 3 = 1_2\cap 3^* = \{0\}$, Appendix \ref{octonion appendix}. As a result, all elements of $\mathfrak{su}$(3) are linearly independent from the spaces of maps:
\beqn
\label{3map}
F_3:3\to 1_2\qquad\text{ and }\qquad F_{3^*}:3^*\to 1_2.
\eeqn
Additionally $\mathfrak{su}$(2) is trivially linearly independent from $F_3$ and $F_{3^*}$.

Now because we consider only subspaces that transform irreducibly under the gauge groups, this implies either $J\cap 2=\{0\}$ or $J\cap2=2$. Then as clearly we must have $1_2\subset J$, this immediately implies we must have $1_2=J$. To see this note that $J\cap 3={0}$ implies Dim$(J)\leq5$; thus the direct sum of $1_2$ with either $2$ or $2^*$ would imply a $J$ of dimension 6, which must have a non-trivial overlap with our triplet of SU(3). Therefore, the only maps transforming in the fundamental representation of SU(3) are the set of maps (\ref{3map}).

Next, it is clear that all the maps $:1\oplus 1^*\to V$, which are singlets of SU(3), are linearly independent from the combined space of maps $F_3\oplus F_{3^*}\oplus \mathfrak{su}(2)\oplus\mathfrak{su}(3)$, Appendix \ref{octonion appendix}. We further decompose this 16 $\mathds{C}$-dimensional space of maps into the two subspaces

\beqn
\label{1map}
F_1:1\to V\qquad\text{ and }\qquad F_{1^*}:1^*\to V.
\eeqn
The maps $F_1$, $F_{1^*}$, $F_3$, and $F_{3^*}$ together constitute all subspaces of M($V$) transforming as either singlets or under fundamental representations of SU(2)$\times$SU(3).

\subsection{The Non-Fundamental Irreducible Representations}
\label{nonfundreps}

Together the set of maps $\mathfrak{su}$(2), $\mathfrak{su}$(3), (\ref{3map}), and (\ref{1map}) span a 51 $\mathds{C}$-dimensional subspace of the 64 $\mathds{C}$-dimensional space of maps M($V$). Therefore there exists an additional set of 13 linearly independent maps not encapsulated by the irreducible representation spaces thus far described. Of these maps we know that none can be of the form $:3\to V'$ for any subset $V'\subset V$. However, since the only space of maps not-yet fully spanned is a subset of the maps
\beqn
\label{remainingspace}
:3\oplus 3^*\to2\oplus2^*
\eeqn
we are forced to consider spaces of maps acting on $3\oplus 3^*$ in such a way that the maps are invariant under complex conjugation and none are of the form $:3\to V'$. So we need maps which simultaneously act on $3$ and $3^*$, which in turn implies we need spaces invariant under complex conjugation.

There exists an obvious choice of maps which satisfy these criteria and map $3\oplus 3^*$ to $2\oplus 2^*$. These spaces will be labelled as $H_1$ and $H_2$ respectively, and are defined as
\begin{align}
\label{higgsrep1}
H_1:=\text{Span}_{\mathds{R}}\Big\{h\in\text{M}(V)\| h:3\to2\text{,  }h:3^*\to2^*;\text{ s.t. }  h(3)^*= \pm h(3^*),
\Big\}
\\\label{higgsrep2}
H_2:=\text{Span}_{\mathds{R}}\Big\{h\in\text{M}(V)\| h:3\to2^*\text{,  }h:3^*\to2;\text{ s.t. }  h(3)^*= \pm h(3^*).
\Big\}
\end{align}
Both $H_1$ and $H_2$ are 12-dimensional vector spaces defined over $\mathds{R}$.

Note that as these subspaces transform irreducibly they are consistent with the particle representation, condition \ref{cond1}. Even so, the non-fundamental and non-adjoint transformations of the spaces $H_i$ do not, at first glance, correspond to any types of particle transformations seen in the Standard Model. As we will see, in section \ref{higgs-like}, the non-fundamental representation spaces of $H_1$ and $H_2$ can be expressed as unique sums of elements transforming under the fundamental representations of SU(2) and SU(3), in this way the choice of $\pm$ will also become arbitrary. These elements naturally inherit properties matching those of the Standard Model Higgs-doublet described in section \ref{smhiggsrep}.

\subsection{Remaining Particle Content}

The maps defined in sections \ref{nonabrep}, \ref{fundandsingreps}, and \ref{nonfundreps} together span a 63-dimensional complex subspace of M($V$). Thus we only lack one particle representation for a full direct sum decomposition of M($V$), naturally this map must be invariant under any transformations of SU(2)$\times$SU(3). The only map we have not considered so far is the map which acts on the entire $3\oplus3^*$, is its own complex conjugate, and is proportional to the identity element when restricted to only $3$ or $3^*$. The map we are describing is exactly the map which would take SU(3) to U(3). Lets denote this map by $Y$. Together our sets of maps span the full direct sum decomposition of M($V$) as:
\beqn
\label{mapdecomposition}
\text{M}(V) =  \mathfrak{su}\text{(2)}\oplus\mathfrak{su}\text{(3)}\oplus F_3 \oplus F_{3^*} \oplus F_1 \oplus F_{1^*} \oplus H_1\oplus H_2\oplus Y.
\eeqn
That is, the algebra of maps on $V$ can be fully decomposed into a direct sum of irreducible representation spaces of SU(3)$\times$SU(2).

It is worth commenting that the map $Y$ is not uniquely defined. Indeed, we could extend this map to act on the entire $V$ in such a way that $Y$ is also proportional to identity when restricted to act only on $1$ or $1^*$. In this case $Y$ becomes interpretable as a U(1) generator preserving the decomposition structure (\ref{su3structure}), yet with relative charges not fixed. That is $Y$ may yield different eigenvalues when acting on $3$ vs. $1$. However, even though the map $Y$ may not be unique, the representation it generates is uniquely one of a U(1) charge generator, as it must preserve the decomposition (\ref{su3structure}) and commute with the $\mathfrak{su}$(3) generators.

Fortunately, the uniqueness of the form of $Y$ is irrelevant for the particle interpretation of the decomposition of M($V$). This is easily seen by noting that any different choice of map $Y'$ affects the decomposition only by a change $Y\to Y'$ in (\ref{mapdecomposition}), and preserves the U(1) particle interpretation. Further, we are here only dealing with representations of groups. In a gauge theory, such as the Standard Model, consistent U(1) charge distributions are restricted by gauge anomaly cancellation, Refs.~\cite{Geng:1988pr, Minahan:1989vd, Babu:1989ex}.

\section{Comments on Decomposition}
\label{comments}

There are three distinct types of spaces in the decomposition (\ref{mapdecomposition}). We will refer to these different types of subspaces, suggestively, as: gauge-like spaces, F-spaces, and Higgs-like spaces. Of course ``Higgs-like" does not imply a particle with similar dynamics to the Higgs doublet, as we do not have any notion of dynamics in our setup. Rather these suggestive names characterize only properties of their representations, in comparison with representations of particles in the Standard Model. Indeed, the gauge-like spaces will be the Lie algebras themselves, as gauge fields transform in the adjoint representation. The F-spaces transform either as a singlet or in the $N$ representation of SU($N$), hence the suggestive ``F" to denote the types of transformations seen by fermions in the Standard Model. Finally the Higgs-like spaces mirror the representation properties of the Standard Model Higgs laid out in section \ref{smhiggsrep}. The following subsections elaborates further on these types of spaces, and their comparison with representations in the Standard Model.

\subsection{Gauge-like Subspaces}
\label{gauge-like}

We have the maps $\mathfrak{su}$(2), $\mathfrak{su}$(3), and Y which are maps that preserve certain irreducible representations of subgroups of $G_2$. Now while our non-abelian Lie algebra maps are generated by subgroups of $G_2$ action on our space $\mathds{C}^8$, the same is not true for our $Y$ generator. Returning to condition \ref{cond1}, this only demands that all maps are between subspaces that transform irreducibly under restricted subsets of $G$. This means that there can be no other non-abelian group structure other than our SU(2)$\times$SU(3) structure, as any other non-abelian group structure would conflict with the decompositions of $V$. However, this statement does not affect the addition of U(1) symmetries, which leave the decompositions (\ref{su3structure}) and (\ref{su2structure}) invariant. Therefore we find the existence of the U(1) generator $Y$ to be perfectly consistent with the principles of construction laid out in section \ref{axiomatisation}. Further as the transformations of the U(1) generator is simply $Y\to Y$, and our $\mathfrak{su}(N)$ transform in the adjoint representation we have subspaces transforming exactly like gauge fields under their respective gauge groups.

\subsection{The $F$ - Subspaces}
\label{Freps}

To better understand the representations contained in the decomposition (\ref{mapdecomposition}), we write out the set of gauge representations contained in the  $F$ - subspaces as
\begin{align}
\label{fundsu3reps}
F_3\quad\ &\text{contains}\quad 4\times(3,1)\\
F_1\quad\ &\text{contains}\quad (1,2) + (1,2^*) + 4\times(1,1)
\end{align}
with the representations of the complex conjugate subspaces following immediately. Note that in (\ref{fundsu3reps}) we have no particles simultaneously transforming under SU(3) and SU(2). Thus we have that certain gauge combinations, such as $(3,1)$, are ``favoured" over other gauge combinations, such as $(3,2)$. This is precisely a variant of the general phenomenon of a non-uniform realization of gauge representation, introduced in section \ref{smhiggsrep}.

Now the fact that not all fundamental representations are realized equally in the construction should come as no surprise. Indeed, if we considered maps between (\ref{su3structure}) and (\ref{su2structure}) without including the adjoint Lie algebra representations, then we would have all the representations
\beqn
(1,1),\,(1,2),\,(1,2^*),\,(3,1),\,(3^*,1),\,(3,2),\,(3,2^*),\,(3^*,2),\text{ and }(3^*,2^*),
\eeqn
appear in the decomposition of M($V$). Therefore this non-uniform realization of gauge representations is a direct consequence of representing all particles, both those transforming in the adjoint and fundamental representations, as elements of the same space.

Of course, the set of irreducible representations derived in this construction do not resemble those of the Standard Model, other than both sets of representations transforming under some SU(2)$\times$SU(3); nor was there any attempt to reproduce Standard Model particle content. Indeed, since we only have two non-abelian gauge groups in the construction we cannot get the same level of complexity in gauge discrimination as in the Standard Model where there are three non-abelian groups to consider. However, the purpose of this paper is to illustrate explicitly that the general construction discussed in section \ref{motivation} can indeed yield a fixed particle content for a gauge theory.

\subsection{Higgs-like Subspaces}
\label{higgs-like}

The remaining spaces of our decomposition (\ref{mapdecomposition}) are then $H_1$ and $H_2$. These maps are denoted by the letter $H$, because they mirror properties of the Standard Model Higgs doublet detailed in section \ref{smhiggsrep}. To show this we need only focus on one of the spaces, lets say $H_1$, as the procedure is identical for $H_2$.

According to the definition of $H_1$ in (\ref{higgsrep1}) elements of this space are maps $h_1$ on $V$ which map $3$ to $2$ and $3^*$ to $2^*$. Now consider a map $\Phi_1$ which acts on the subspace $3$ such that $\Phi_1(3)\equiv h_1(3)$, and vanishes on the rest of $V$, i.e. $\Phi:V\backslash 3 \to\{0\}$. To fully describe a map $h_1\in H_1$, we would also need a map $\Theta_1$ which acts on $3^*$ in such a way that $\Theta_1(3^*)\equiv h_1(3^*)$. Then we have that $h_1=\Phi_1\pm\Theta_1$. However, the maps $\Phi_1$ and $\Theta_1$ are not independent maps. Indeed from the definition of $H_1$, in (\ref{higgsrep1}), we see that $\Theta_1(3^*)\equiv h_1(3^*) = \pm[h_1(3)]^* \equiv \pm[\Phi_1(3)]^*$. Therefore we may express any map $h_1\in H_1$ as the sum
\beqn
h_1= \Phi_1\pm\Phi_1^*,
\eeqn
where $\Phi_1$ is a 6-dimensional complex subspace of M($V$) transforming in the fundamental representation of SU(2) and SU(3). 

In the Standard Model, the Higgs doublet is such that it forms gauge-invariant terms when contracting with a $(3,2)$ left handed spinor and a $(3,1)$ right handed spinor. As outlined in section \ref{smhiggsrep}, for the Yukawa interactions to yield masses to both up-type and down-type quarks after spontaneous symmetry breaking this requires the use of both the Higgs-doublet and its complex conjugate.

In the construction presented here, we can see that a map $h_1\in H_1$ can form invariants with representations of the SU(3)$\times$SU(2) group by contracting maps in the $(3,1)$ representation with maps in the $(1,2)$ representation. Additionally, using the same map $h_1\in H_1$, we can also form invariants by contracting maps in the $(3^*,1)$ representation with maps in the $(1,2^*)$ representation. Equivalently we can describe these contractions with $h_1$ via the map $\Phi_1$ and its complex conjugate. In this case we use $\Phi_1$ to form invariants when contracting maps in the $(3,1)$ representation with maps in the $(1,2)$. Similarly, instead of $h_1$ we can equivalently use $\Phi_1^*$ to form invariants when contracting maps in the $(3^*,1)$ representation with maps in the $(1,2^*)$.

Thus viewing the subspaces $H_i$ as describing particles $\Phi_i$, we can form all relevant invariants if and only if we use both the map $\Phi_i$ and its complex conjugate $\Phi_i^*$. Consequently, the spaces $H_i$ describe ``Higgs-like" representations $\Phi_i$, mirroring properties highlighted in section \ref{smhiggsrep} for the Standard Model Higgs doublet. A crucial property we cannot comment on is the Lorentz representation of the Higgs doublet, as we have no Lorentz group in our setup.%
\footnote{Only a scalar field may obtain a vacuum expectation value without spontaneously breaking the Lorentz symmetry of the vacuum. Since our set-up does not contain Lorentz symmetries, realizing this property of the Higgs field is beyond our simple construction.} 
Note that when working directly with $\Phi_i$ the choice of $\pm$  in (\ref{higgsrep1}) is arbitrary, as an overall sign can be absorbed into the coupling coefficient in the action.

\section{Vector Spaces and Algebras}
\label{vectoralgpurespin}

Note that the only relevant properties of $(V,G)$ for our construction are the irreducible representation of $V$ under $G$ and its subgroups. Therefore the decomposition of M($V$) cannot distinguish between a pair $(V,G)$ and an algebra $(A,\cdot)$ for which Dim$(A)=\text{Dim}(V)$ and the inner automorphism group of $A$ is equal to $G$. Consequently maps on the structure $(\mathds{C}^8,G_2)$ are equivalent to maps on the complexification of the Octonions, $\mathds{O}$, whose inner automorphism group is exactly $G_2$.

Using an algebra instead of a vector space, while seemingly equivalent, might yield deeper insights into the construction. Indeed the appearance of the SU(2) and SU(3) subgroups of $G_2$ are related to the preservation of invariants within the Octonionic algebra. Specifically any SU(3)$\subset G_2$ can be uniquely defined as the subgroup of $G_2$ that preserves a complex structure on $\mathds{O}$. Similarly, SU(2)$\subset G_2$ is uniquely defined as the subgroup of $G_2$ that preserves some Quaternionic subspace of $G_2$, for more details see Appendix \ref{octonion appendix}. Invariants are of crucial importance to particle physics, as all experimentally established theories of particle physics rely on the use of invariants. Therefore working with an algebra and its automorphism group $(A,\text{Aut}(A))$, as opposed to some pair $(V,G)$, may offer selection rules for the decompositions of $A$, even when $G\equiv \text{Aut}(A)$. Thus the route to uniqueness of the construction may require the use of algebras instead of vector spaces.

\section{Generalizations}
\label{generalization}

We have presented a simple model, as a proof of concept, to show how a set of particle representations can arise as endofunctions on an eight $\mathds{C}$-dimensional vector space, equipped with the symmetry group $G_2$. The general construction relies fundamentally on the choice of pair $(V,G)$ and the set of conditions outlined in section \ref{axiomatisation}. As such, there are several avenues to explore in order to generalize the construction provided here. Below we outline how, and why, one may wish to modify the presented construction.

\subsection{Choice of $(V,G)$}
\label{VGchoice}

The most clear extension of this construction would be to consider different spaces $V$ and symmetry groups $G$. Such a generalization is essential if any connection is to be made with modern particle theory, primarily because $G_2$ does not contain the SL(2,$\mathds{C}$) gauge group necessary to obtain Lorentz representations. The potential of the construction presented here goes just reproducing Standard Model particle representations. Indeed, as the construction imposes restrictions on the appearance of gauge structure, extensions of this procedure could be used to generate representations with both chiral discrimination and the appearance of generations. Using the language of modern gauge theorists, we would say we found a SU(2) discriminating behaviour in our simple construction involving only the subgroups SU(2) and SU(3). It would therefore be interesting to investigate whether there exists a pair $(V,G)$ that can describe a SL(2,$\mathds{C}$)$\times$SU(3) chiral discrimination of SU(2).%
\footnote{Note that while our construction considers maps between two irreducible representation spaces, it is not limited to a choice of only two non-abelian groups. For example a SL$(2,\mathds{C})\times G_2$ group structure on some space $V$ would yield subspaces of M($V$) transforming under representations of SL(2,$\mathds{C}$)$\times$SU(3)$\times$SU(2).}

Further, the choice of $(V,G)$ could be related to the existence of ``generations", i.e. multiple copies of the same particle representations. In the construction presented here, the simplest way that generations can appear if a subgroup $K\subset G$ is such that there exists a set of distinct subspaces $\{S_i\}_{i=1}^n$, where all $S_i$ transform under the same irreducible representation of $K$. Explicitly, let $R$ be some irreducible representation of $Z\subset G$, where $Z{\displaystyle \not \subset }K$. Then the maps $F_i:R\to S_i$ form a set of $n$ linearly independent sets of maps transforming in the same representations, i.e. n distinct particle generations.

\subsection{Choice of Construction Principles}
\label{principlechoice}

In section \ref{axiomatisation} we outlined a set of conditions, describing construction principles, that were essential in deriving the decomposition (\ref{mapdecomposition}) of particle representations. This naturally begs the question of whether this choice of conditions is necessary and/or unique. It is clear that conditions \ref{cond1} and \ref{cond2}, or some similar form of these conditions, are essential. For example, condition \ref{cond1} describes how to induce structure on M($V$) from the space $V$ and its symmetry group $G$. Such a condition is clearly necessary as we have to assign representations to subspaces of M($V$). Similarly, \ref{cond2} is the minimal condition for a direct-sum decomposition of M($V$) into particle representations. In other words, while \ref{cond1} determines the types of representation spaces of M($V$), condition \ref{cond2} specifies how to relate these representation spaces to a particle content.

Next, all gauge theories containing Lorentz spinors require spinor antiparticles as a consequence of charge conservation.%
\footnote{This is directly related to the interpretation of the hermitian-conjugate, or dual, spinors as antiparticles.}
Since we are seeking a mechanism by which to generate gauge theories, it is clear that we require a decomposition of M($V$) which ensures the existence of representations which may be used to construct a gauge-invariant field theory. This was exactly the purpose of condition \ref{cond3}, which required the existence of antiparticle representations for each particle representation. However, the specific form of this condition could be generalized in many ways and still yield the particle-antiparticle relation relevant for $(\mathds{C}^8,G_2)$.

Indeed, complex conjugation is the only outer automorphism of our setup. Thus we could equivalently have rephrased condition \ref{cond3} as invariance of the decomposition of M($V$) under the outer automorphism group of $V$. In this case condition \ref{cond3} may not reduce to a particle-antiparticle relation for different choices of $(V,G)$, or may imply further conditions on the particle content than only a particle-antiparticle relation. For example, an outer automorphism group which contains the cyclic group could then yield the existence of multiple generations of particles.%
\footnote{For example, a cyclic group of order $3$ as an outer automorphism could keep SU(3) representation spaces invariant, but cycle a triplet of SU(2) representation spaces. This is exactly the case for the cyclic group $C_3$ acting on the Octonions $\mathds{O}$, where $C_3 \equiv \text{Aut}(\mathds{O})\slash G_2$. For a more detailed analysis on the outer automorphism group of the octonions see Ref.~\cite{Furey:2016ovx}.}
Alternatively it is not clear that a third condition is always needed to derive a direct sum decomposition. A natural question is whether there exists choices of $(V,G)$ for which a particle-antiparticle relationship appears as the only decomposition consistent with conditions \ref{cond1} and \ref{cond2}, or some variant thereof.

\noindent{In general these questions may be concisely be expressed as:}
\begin{itemize}
	\item[] Does there exist a unique set of conditions which, for any  $(V,G)$, ensures the existence of a direct-sum decomposition of M($V$) into subspaces which transform under irreducible representations of non-trivial subgroups of $G$?
		\subitem -If no: For what pairs may such a decomposition be found?
		\subitem -For $(V,G)$ admitting such a decomposition: What conditions are needed to ensure the decomposition's uniqueness?
\end{itemize}

\section{Distinct Features}
\label{predictions}

While many of the results that follow depend on the set of choices outlined in sections \ref{VGchoice} and \ref{principlechoice}, there are some consequences for particle representations that are universal for the general construction method outlined here.

One important consequence is that to reproduce the Standard Model interactions requires additional particle content. This can be seen in a number of ways, but most clearly we can note that the set of maps must span a matrix algebra. This implies that to span the matrix space the total number of components needed to describe all particle fields must equal to a perfect square. This is clearly not the case for the Standard Model particle content, Ref.~\cite{Gording:2019srz}. However, one can make even stronger predictions.

In fact it follows that to describe the entire Standard Model particle content via the proposed construction minimally requires the inclusion of additional scalar degrees of freedom as well as another gauge group $G'$, containing a U(1) subgroup.%
\footnote{It is possible to obtain the Standard Model gauge structures without the addition of another U(1) symmetry. However, this is only possible if one considers a very large space $V$, such that both the appearance of the Standard Model structures becomes trivial and the Standard Model particle content becomes a minor subset of the full particle spectrum. We here outline only the approach requiring minimal new particle content.}
To see this, consider maps from an SU(3) decomposition of some space $V$, where we only have one triplet and one anti-triplet in the decomposition. Then to yield different charges to quarks and anti-quarks we must have U(1) transformations on our SU(3) decomposition, with opposite eigenvalues (i.e. charges) when acting on $3$ and $3^*$. Lets call this U(1) group the group U(1)$_A$, which acts only on the SU(3) decomposition of $V$. The U(1)$_A$ transformations will however yield the same charges for up- and down-type quarks, as the different types of quarks have the same SU(3) structure. That is both up- and down-type quarks must map from $3\subset V$, thus yielding the same U(1)$_A$ charges. Therefore, to yield distinct charge assignments we require a map from the SU(3) decomposition to some other decomposition of $V$ which carries a different U(1) symmetry, lets refer to this group as U(1)$_B$.

\begin{figure}
	\centering
	\includegraphics[scale=0.5]{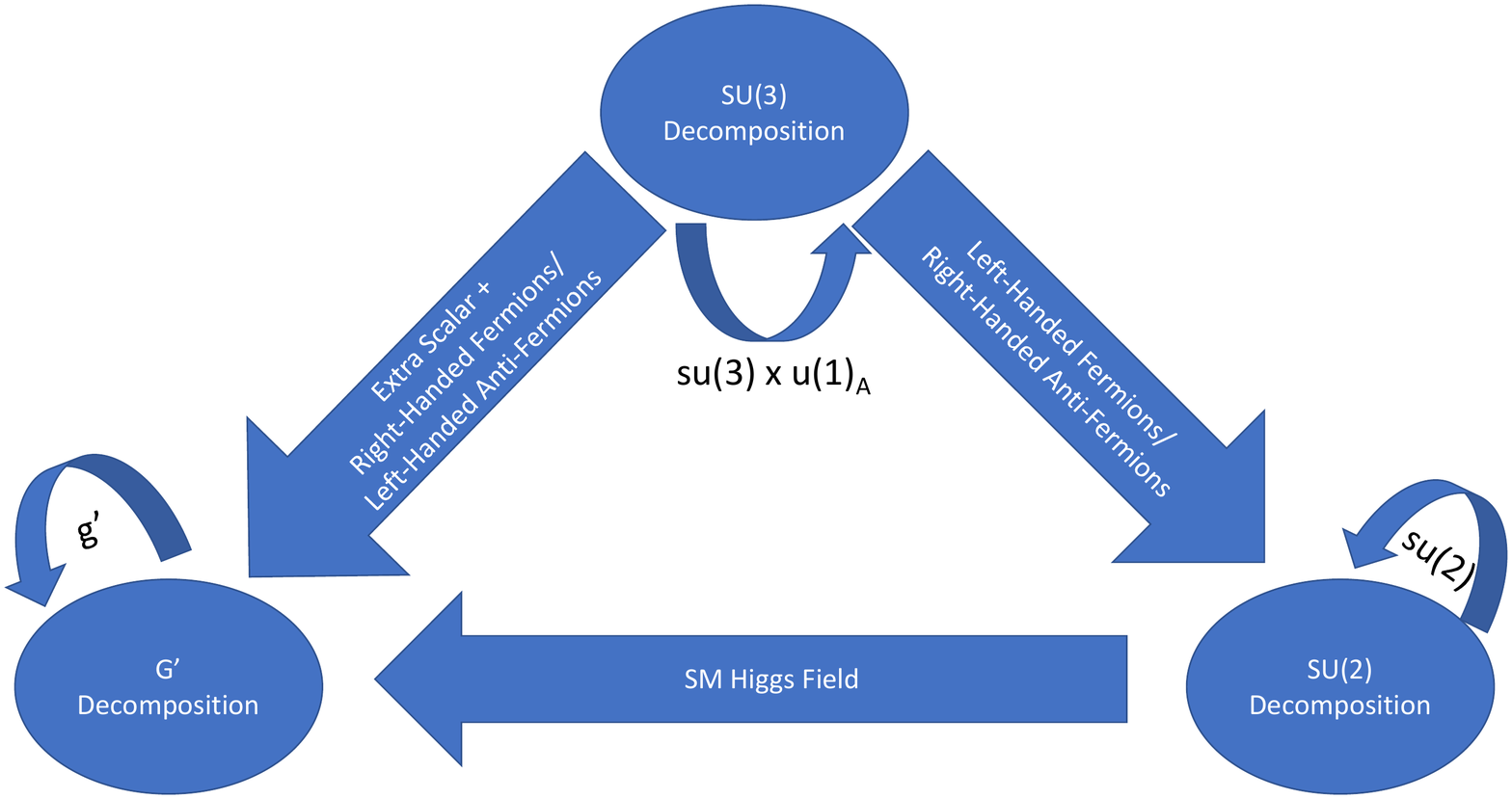}
	\caption{Pictographic illustration of a minimal approach to generating different charges for up- and down-type quarks. Here $\mathfrak{g}'$ is the Lie algebra of $G'$.}
	\label{fig: fitwithhiggs}
\end{figure}

Now, for left handed fermions the U(1)$_B$ symmetry needed to yield correct charges comes from the weak-isospin generator during spontaneous symmetry breaking.%
\footnote{That is before spontaneous symmetry breaking there is no charge distinction between left handed up- and down-type quarks. We need an additional generator from the electroweak interactions to be able to generate different charges. This additional generator then takes the role of U(1)$_B$ for left-handed fermions.}
However, in the Standard Model there is no additional symmetry group for right handed fermions for which we may find a U(1)$_B$ generator to yield distinct up- and down-type charges. Thus the construction outlined in this paper demands at minimally an additional U(1) symmetry.%
\footnote{Note that the additional symmetry group does not need to be exactly a U(1) group. Rather, the only requirement is that we have some group $G'$ for which $\text{U(1)}_B\subset G'$.}
We now have two independent U(1) symmetries, but only have one U(1) symmetry in the Standard Model. Therefore we require an additional scalar field to acquire a vacuum expectation value and spontaneously break the U(1)$_A\times$U(1)$_B$ symmetry down to the weak hypercharge of the Standard Model. We illustrate an example of such a set-up in \textbf{Figure \ref{fig: fitwithhiggs}}.

The requirement of this additional particle content, and symmetry structure, has some immediate implications. First, in the minimal set up there would only be one triplet, $3$, and one anti-triplet, $3^*$, in the SU(3) decomposition. Therefore U(1)$_A$ would yield the same charge for each particle generation. This would then imply the U(1)$_B$ charges must be the same for each generation of fermions. This follows from the weak hypercharge of the Standard Model acting identically on all generations. This is consistent with observations at current energies regarding Lepton-Universality, Ref.~\cite{Aaij:2019wad}. Second, depending on the form of the action, after spontaneous symmetry breaking the resultant massive scalar field could be very light and interact only weakly with the broken U(1) generator. As a result, under such a set-up, one could have a particle theory containing a weakly-interacting massive scalar field, i.e. a ``WIMP". Such fields have interesting cosmological applications \cite{Jungman:1995df,ArkaniHamed:2008qn}.

The above analysis makes it clear that while the procedure for constructing particle representations presented in this paper is very general, any resultant theories come with distinct and falsifiable properties. In the current set up there are no specific predictions, as we have no theory from which to deduce relative interaction strengths or mass scales. Nevertheless, the distinct features that are associated with the construction of gauge theories as endofunctions on an underlying space implies that any adaptation to Standard Model physics will yield testable predictions.

\section{Comparison with Non-commutative Geometry}
\label{noncommutativegeometry}

The familiar reader may at this point have noticed that our construction bears many similarities to non-commutative geometry. These features arose independently as necessary consequences of our approach. We will here discuss some similarities and advantages of each approach. We note, naturally, that the field of noncommutative geometry is much more developed, and as such there are many features of which simply cannot be compared to our approach here.

A fundamental aspect of both these approaches to understanding the appearance of specific particle representations is the use of a set of specific principles from which to derive the particle representations. In our construction we have referred to these principles as ``construction principles", while non-commutative geometry employ a set of axioms, Ref.~\cite{Connes:1996gi}. The purpose of such principles is to remove arbitrary choice from the construction of particle representations, which is essential when creating an approach to generating unique particle content. As already mentioned in section \ref{axiomatisation}, the construction principles of this paper are not enough to ensure uniqueness of the decomposition (\ref{mapdecomposition}). However, for our purposes this was intentional, as we found the presentation of a full set of axioms may contain biases specific to the particular example of ($\mathds{C}^8$, $G_2$) discussed here. As such it would be interesting to compare a full set of axioms appearing from a further development of our approach to those of non-commutative geometry. 

The way in which particle representations appear in these two approaches is fundamentally distinct. This is because non-commutative geometry treats fermions and gauge bosons as belonging to distinct spaces. Specifically, fermions belong to some Hilbert space of states, and bosons are operators acting on this space. As such, non-commutative geometry attempts to unify independently the bosonic and fermionic sectors of the Standard Model. This differs from our approach, where both fermions and bosons appear as elements of the same space. As such our approach seeks to provide one further level of unification, where all particles live in the same space. The difference arises as instead of working with a Hilbert space of fermion states and bosonic operators acting on this space, we are working with an algebra, i.e. M($V$). This is possible because the operations describing gauge transformations of fermions in M($V$) can themselves be represented as elements of the algebra, and thus bosons acting on fermions is simply the algebra acting on itself.%
\footnote{Technically we cannot comment on bosons versus fermions, as our approach does not yet contain any spacetime representations and thus we cannot define spin eigenstates. Instead, we here refer to fermions as the particles in our $F-$ spaces and bosons as the particles in the remaining subspaces of (\ref{mapdecomposition}). This is consistent with our identification of ``gauge-like" and ``Higgs-like" representations.}
Describing both bosons and fermions as elements of the same space has the additional benefit that the explicit appearance of three generations of particles may be explained in our approach as a consequence of linear independence. Non-commutative geometry does not yet have any explanation for why the fermionic representations appear in multiplicity of three. Ref~\cite{Chamseddine:2012sw}.

Another important point of comparison is the explicit appearance of both particle and anti-particle representations, Ref.~\cite{Connes:1996gi,Chamseddine:1996zu}. These representations are related by an anti-linear isometry, denoted by $J$, and because fermions and bosons occupy different spaces in non-commutative geometry there are two separate relations involving $J$. One relation specifies how fermions are related to the anti-fermions. The other relation is the (anti-)commutation%
\footnote{The relation will either be commutation or anti-commutation depending on the dimension of the spacetime manifold, as detailed in Ref.~\cite{Chamseddine:1996zu}.}
of this operator w.r.t. the gauge covariant derivative, and implies that bosons are their own anti-particles. Such an operator is also defined within our approach, namely the operation of complex conjugation. However, since in our approach both bosons and fermions lie in the same space, we need only one formula to specify both how fermions are related to anti-fermions and implies that bosons are their own anti-particles. The appearance of this complex conjugation operator was stipulated by condition \ref{cond3}.

The last point of comparison we will discuss is that in both approaches there are implied additional scalar degrees of freedom necessary to match the constructions with the Standard Model particle representations. Furthermore in both the approaches these additional scalar components are implied by the construction, Ref.~\cite{Chamseddine:2012sw}.

There are many direct similarities between non-commutative geometry and the construction proposed here. However, there are also many features of non-commutative geometry for which we cannot compare to our approach simply because our approach lacks sufficient development. For example, non-commutative geometry presents a unification of gauge and spacetime symmetries, Ref.~\cite{Chamseddine:1996zu,Chamseddine:2014uma}, into one framework for deriving bosonic fields as ``internal" fluctuations of a metric and the gravitation fields as ``external" fluctuations. As our approach has not yet considered spatial symmetries this is clearly not possible. Further, non-commutative geometry has a method for constructing a unique (up to gauge couplings) action through considering the spectrum of the gauge covariant derivative, and its action on the fermion states, Ref.~\cite{Connes:1996gi,Chamseddine:1996zu}. Our work has so far only been focused with the identification of particle representations, and thus no consideration has yet been given to the development of a mechanism for generating an action. 

Clearly non-commutative geometry stands on a much more solid footing than the construction we have presented here, as a consequence of being a much more developed approach. Even so, we emphasize that there are distinct advantages to having one unifying space which describes both fermionic and bosonic representations. As such this construction warrants further development, as its formulation has the potential to answer certain questions which have not yet been answered within the context of non-commutative geometry. It seems likely that further development of our proposed construction could be assisted by borrowing concepts that have already been developed for non-commutative geometry, as the two approaches have enough similarities to make certain concepts transferable.

\section{Conclusion}
\label{conclusion}

We have presented a construction which takes a vector space $V$ together with a symmetry group $G$, and returns a complete and finite set of particle representations. As a proof of concept we picked the pair $(\mathds{C}^8,G_2)$, due to the appearance of the non-abelian Standard Model gauge groups SU(2) and SU(3) as independent subgroups of $G_2$. To ensure a consistent approach we formulate a set of construction principles to guide the decomposition of the space of maps M($V$). With these principles we derive a decomposition of M($V$) into maps between the irreducible representation spaces of SU(2) and SU(3). Further, we find that not all representations appear equally in the decomposition of M($V$) and we discuss how this generalizes the phenomenon of SU(2) only acting on left-handed fermions in the Standard Model.

Even in our simple example, we find that the set of all maps in M($V$) may be grouped into three categories. These are referred to as the gauge-like, $F$-, and Higgs-like subspaces of M($V$) and detailed in sections \ref{gauge-like} - \ref{higgs-like} respectively. These are so named as they seem to correspond to the three different types of Standard Model particle representations. We can not, at this early stage, say for certain whether the appearance of Higgs-like subspaces is a consequence of the construction or choice of the pair ($\mathds{C}^8,G_2$). However, it is clear that the construction will always yield gauge-like and $F$- subspaces, as the construction requires both endofunctions on and maps between decompositions of the underlying space $V$. There is also good reason to believe the appearance of gauge-like subspaces results in the existence of Higgs-like subspaces to ensure linear independence, as was the case for our choice of ($\mathds{C}^8,G_2$). We leave further explorations of this type to future work.

The purpose of the construction is to provide a framework to generate a finite and fixed set of particle representations for use in a gauge theory construction. The construction contains many points of comparison to non-commutative geometry, and we discuss some key similarities and differences between the two approaches in section \ref{noncommutativegeometry}. To our knowledge, this paper presents a unique approach to constructing particle representations for a gauge theory. We did not attempt the ambitious task of reproducing the Standard Model representations, but instead provide the proof of concept for a much simpler setup. Therefore, it is clear that to establish a direct connection to modern particle physics more development is needed. In particular the questions raised in the discussion of section \ref{principlechoice} must be answered and clarified.

Further, for the construction to be capable of producing representations of particles in spacetime, Lorentz transformations must be incorporated into the construction mechanism. For consistency this implies one set of construction principles from which both the Lorentz and gauge representations follow naturally. Such general construction principles would describe a unified origin of both the ``internal" gauge and ``external" Lorentz symmetries. As such it may lead to insights in how to extend the gauging of the group $G_{\text{SM}}$ to include the spatial representations of particles.

Finally, as the theory cannot reproduce only the Standard Model particle content, any application of the proposed construction to Standard Model physics will yield testable predictions. This in turn implies any resultant theories are falsifiable, showing that the gauge-theory construction herein proposed is not arbitrary and would yield real, observable consequences.

\vspace{30pt}
\paragraph{Acknowledgements.} This work is supported by a grant from the Max-Planck-Society.

\newpage

\appendix

\section{:$\qquad$ SU(2) and SU(3) Decompositions of $\mathds{C}\otimes\mathds{O}$}
\label{octonion appendix}

We here prove statements given in section \ref{derivation} about the irreducible representation spaces of ($\mathds{C}^8,G_2$) under SU(2) and SU(3). To do so we will be using that as far as endofunctions on $\mathds{C}^8$ are concerned, we may equivalently work with the algebra $\mathds{C}\otimes\mathds{O}$ and the automorphism group $G_2$. Firstly, we note that any SU(3)$\subset G_2$ appears uniquely as the subgroup of $G_2$ that preserves some unit imaginary element of $\mathds{O}$. This is naturally in addition to the identity element which is invariant under $G_2$. As unit imaginary elements square to $-1$, SU(3)$\subset G_2$ is the group which preserves some unique complex structure on $\mathds{O}$. We can extend this idea to preserve two unit imaginary elements of $\mathds{O}$. However, leaving any two unit imaginary elements $a,b\in\mathds{O}$ invariant implies that their product $c:=ab$ must also be left invariant, where $c\in\mathds{O}$ must also be an imaginary unit element. Therefore, the next smallest subgroup of $G_2$ will be such that it leaves a Quaternionic subalgebra Span$\{1,a,b,c\}\subset\mathds{O}$ invariant. This group is precisely SU(2)$\subset G_2$.

Note that if one wishes to extend this construction further one immediately encounters that the only subgroup of $G_2$ which keeps four unit imaginary Octonionic elements invariant is the trivial group $\{\text{I}\}\subset G_2$, where $I$ is the identity element of $G_2$. This is because given a Quaternionic triplet $\{a,b,c\}$ and some other unit imaginary $d$, preserving all these elements also implies preserving their products. Or in other words, preserving all elements $\{a,b,c,d,(ad),(bd),(cd)\}$, where $ad$ denotes the Octonionic product of the elements $a$ and $d$. However, this set spans all unit imaginary elements of $\mathds{O}$, and thus we are leaving all of $\mathds{O}$ invariant. Therefore, the only unique non-trivial subalgebras of $G_2$ related to invariant subspaces of $\mathds{O}$ are SU(3) and SU(2).%
\footnote{Note that we could also define U(1) transformations which leave elements of $\mathds{O}$ invariant, however these U(1) groups would exist as subgroups of either SU(3) or SU(2).}

Now, the three distinct $\mathds{C}$-subalgebras of a Quaternionic subalgebra correspond precisely to the fact that each SU(3) contains three distinct SU(2) subgroups. Therefore, to ensure that SU(2) is not a subgroup of SU(3), we must ensure that the Quaternionic subspace preserved by SU(2) does not contain the complex subspace preserved by SU(3). Let the complex subspace preserved by SU(3) be spanned by the basis elements $\{1,d\}$ and the Quaternionic subspace preserved by SU(2) be spanned by basis elements $\{1,a,b,c\}$. Then it becomes clear that triplet, $3$, and conjugate-triplet, $3^*$, subspaces of $\mathds{C}\otimes\mathds{O}$ are in distinct eigenspaces of $d$. In particular,
\beqn
3 &= \text{Span}\{a-i(da),b-i(db),c-i(dc)\}
\\
3^* &= \text{Span}\{a+i(da),b+i(db),c+i(dc)\}
\eeqn

Similarly the $2$ and $2^*$ subspaces of $\mathds{C}\otimes\mathds{O}$ together span
\beqn
2\oplus2^* = \text{Span}\{d,(da),(db),(dc)\}
\label{octonionic doublet span}
\eeqn
The subspaces $2$ and $2^*$ are then defined as the conjugate pair of eigen-subspaces of (\ref{octonionic doublet span}) under left action of either $a$, $b$, or $c$.%
\footnote{Clearly there is more freedom in how to pick $2$ and $2^*$ than we had freedom in picking our $3$ and $3^*$. However, picking some arbitrary $2$ and $2^*$ pair essentially only corresponds to a basis choice in the SU(2) decomposition of $V$.}
It is easy to verify that irrespective of which $a$, $b$, or $c$ we chose to define $2$ and $2^*$, we have that $3\cap 2=3\cap2^*=\{0\}$. Further $1_2\equiv\text{Span}\{1,a,b,c\}$, which is clearly linearly independent from either $3$ or $3^*$, but not from $3\oplus3^*$.

Finally, we look at the subspace $1\oplus1$ in (\ref{su3structure}). This is precisely the complex subspace spanned by $\{1,d\}$ and is therefore trivially linearly independent from $3$ and $3^*$. However, for the analysis in section \ref{fundandsingreps}, we must show that $1\oplus1$ is linearly independent from $2$ and $2^*$. This is simply done by noting that since $1\subset\mathds{O}$ is trivially linearly independent from either of $2$ and $2^*$, we must only show that $d$ is also linearly independent from the doublets. Now clearly $d\subset 2\oplus2^*$. However, we only care about linear independence from $2$ and $2^*$ individually. Therefore, without loss of generality, let $\{\varepsilon_1,\varepsilon_2,\varepsilon_3\} =\{a,b,c\}$ be such that $\{\varepsilon_i\}$ satisfies the commutation relation of unit Quaternionic basis elements. Then for some arbitrary $\varepsilon_j$ we may define $2$ and $2^*$ as
\beqn
2 &:= \text{Span}\{d+i(d\varepsilon_j),(d\varepsilon_{j+1})+i(d\varepsilon_{j+2})\}
\\
2^* &:= \text{Span}\{d-i(d\varepsilon_j),(d\varepsilon_{j+1})-i(d\varepsilon_{j+2})\}
\eeqn
where the indices on $\varepsilon_j$ are defined up to mod 3, i.e. $\varepsilon_4\equiv\varepsilon_1$. Then, as $d\cap 2=d\cap2^*=\{0\}$, we have shown that $1\oplus1$ in (\ref{su3structure}) is linearly independent from $2$ and $2^*$ individually.

\section{Disclaimer: Incorrect Expression for subspaces $H_1$ and $H_2$}

The identification of the $H$-spaces in section \ref{nonfundreps} is incorrect. The arguments in this section fall short and we have found, to be shown in an upcoming publication, that the spaces $H_1$ and $H_2$ are in fact not linearly independent from the rest of the irreducible representation spaces identified. Instead the linearly independent irreducible representation spaces are:
\begin{align}
\label{correcthiggsrep1}
&H_1^\alpha:=\text{Span}_{\mathds{R}}\Big\{h\in\text{M}(V);\, h=\phi_1+\phi_2\| \phi_1:3\to2\text{,  }\phi_2:3^*\to2^*;  \text{ s.t. }\phi_2=\alpha\phi_1^*
\Big\},
\\\label{correcthiggsrep2}
&H_2^\beta:=\text{Span}_{\mathds{R}}\Big\{h\in\text{M}(V);\, h=\psi_1+\psi_2\| \psi_1:3^*\to2\text{,  }\psi_2:3\to2^*;  \text{ s.t. }\psi_2=\beta\psi_1^*
\Big\}.
\end{align}
These subspaces do satisfy linear independence with the rest of the representation spaces defined in this text. Clearly these new $H$-spaces violate construction principle \ref{cond3}, as these subspaces are not invariant under complex conjugation nor have companion complex conjugate subspaces. We will provide more comments on these new $H$-spaces in our future publication. Ultimately the new subspaces (\ref{correcthiggsrep1}) and (\ref{correcthiggsrep2}) are still interpreted as ``Higgs"-like, with features of interest and comparison to Standard Model representations remaining intact. Therefore this discrepancy does not alter the implications of our work, and is instead a slight modification of the specific form of our induced irreducible representation spaces.


\newpage
\begin{thebibliography}{999}

\bibitem{Sirunyan:2020ztc}
A.~M.~Sirunyan {\it et al.} [CMS Collaboration],
arXiv:2001.10086 [hep-ex].

\bibitem{Azzi:2019yne}
P.~Azzi {\it et al.},
``Report from Working Group 1 : Standard Model Physics at the HL-LHC and HE-LHC,''
CERN Yellow Rep.\ Monogr.\  {\bf 7} (2019) 1
doi:10.23731/CYRM-2019-007.1
[arXiv:1902.04070 [hep-ph]].



\bibitem{Georgi:1974sy}
H.~Georgi and S.~L.~Glashow,
``Unity of All Elementary Particle Forces,''
Phys.\ Rev.\ Lett.\  {\bf 32} (1974) 438.
\bibitem{Weinberg:1974yx}
S.~Weinberg,
``Recent progress in gauge theories of the weak, electromagnetic and strong interactions,''
Rev.\ Mod.\ Phys.\  {\bf 46} (1974) 255
[J.\ Phys.\ Colloq.\  {\bf 34} (1973) no.C1,  45].
\bibitem{Georgi:1974yf}
H.~Georgi, H.~R.~Quinn and S.~Weinberg,
``Hierarchy of Interactions in Unified Gauge Theories,''
Phys.\ Rev.\ Lett.\  {\bf 33} (1974) 451.
\bibitem{Miura:2016krn}
K.~Abe {\it et al.} [Super-Kamiokande Collaboration],
Phys.\ Rev.\ D {\bf 95} (2017) no.1,  012004
[arXiv:1610.03597 [hep-ex]].

\bibitem{Pati:1973uk}
J.~C.~Pati and A.~Salam,
``Unified Lepton-Hadron Symmetry and a Gauge Theory of the Basic Interactions,''
Phys.\ Rev.\ D {\bf 8} (1973) 1240.
doi:10.1103/PhysRevD.8.1240
\bibitem{Pati:1974yy}
J.~C.~Pati and A.~Salam,
``Lepton Number as the Fourth Color,''
Phys.\ Rev.\ D {\bf 10} (1974) 275
Erratum: [Phys.\ Rev.\ D {\bf 11} (1975) 703].
doi:10.1103/PhysRevD.10.275, 10.1103/PhysRevD.11.703.2
\bibitem{Mohapatra:1974hk}
R.~N.~Mohapatra and J.~C.~Pati,
``Left-Right Gauge Symmetry and an Isoconjugate Model of CP Violation,''
Phys.\ Rev.\ D {\bf 11} (1975) 566.
doi:10.1103/PhysRevD.11.566
\bibitem{Georgi:1974my}
H.~Georgi,
``The State of the Art—Gauge Theories,''
AIP Conf.\ Proc.\  {\bf 23} (1975) 575.
doi:10.1063/1.2947450
\bibitem{Georgi:1981vf}
H.~Georgi,
``An Almost Realistic Gauge Hierarchy,''
Phys.\ Lett.\  {\bf 108B} (1982) 283.
doi:10.1016/0370-2693(82)91193-5
\bibitem{Grinstein:1982um}
B.~Grinstein,
``A Supersymmetric SU(5) Gauge Theory with No Gauge Hierarchy Problem,''
Nucl.\ Phys.\ B {\bf 206} (1982) 387.
doi:10.1016/0550-3213(82)90275-9
\bibitem{Masiero:1982fe}
A.~Masiero, D.~V.~Nanopoulos, K.~Tamvakis and T.~Yanagida,
``Naturally Massless Higgs Doublets in Supersymmetric SU(5),''
Phys.\ Lett.\  {\bf 115B} (1982) 380.





\bibitem{Gording:2019srz}
B.~Gording and A.~Schmidt-May,
``The Unified Standard Model,''
arXiv:1909.05641



\bibitem{vonDelft:1998pk}
J.~von Delft and H.~Schoeller,
``Bosonization for beginners: Refermionization for experts,''
Annalen Phys. \textbf{7} (1998), 225-305
[arXiv:cond-mat/9805275 [cond-mat]].

\bibitem{Senechal:1999us}
D.~Senechal,
``An Introduction to bosonization,''
[arXiv:cond-mat/9908262 [cond-mat]].




\bibitem{Burgess:2007zi}
C.~Burgess and G.~Moore, 2006, \textit{The Standard Model: A Primer}, Cambridge University Press, Cambridge UK.



\bibitem{Buchmuller:2006zu}
W.~Buchmuller and C.~Ludeling,
``Field Theory and Standard Model,''
hep-ph/0609174.




\bibitem{GellMann:1962xb}
M.~Gell-Mann,
``Symmetries of baryons and mesons,''
Phys.\ Rev.\  {\bf 125} (1962) 1067.
doi:10.1103/PhysRev.125.1067


\bibitem{CompactLieGroups}
Sepanski, Mark R., 2007, \textit{Compact Lie Groups} (Springer Science+Busines Media LLC, New York)

\bibitem{Geng:1988pr}
C.~Q.~Geng and R.~E.~Marshak,
``Uniqueness of Quark and Lepton Representations in the Standard Model From the Anomalies Viewpoint,''
Phys.\ Rev.\ D {\bf 39} (1989) 693.
\bibitem{Minahan:1989vd}
J.~A.~Minahan, P.~Ramond and R.~C.~Warner,
``A Comment on Anomaly Cancellation in the Standard Model,''
Phys.\ Rev.\ D {\bf 41} (1990) 715.
\bibitem{Babu:1989ex}
K.~S.~Babu and R.~N.~Mohapatra,
``Quantization of Electric Charge From Anomaly Constraints and a Majorana Neutrino,''
Phys.\ Rev.\ D {\bf 41} (1990) 271.




\bibitem{Furey:2015tqa}
C.~Furey,
``Charge quantization from a number operator,''
Phys.\ Lett.\ B {\bf 742} (2015) 195
[arXiv:1603.04078 [hep-th]].
\bibitem{Furey:2016ovx}
C.~Furey,
``Standard model physics from an algebra?,''
arXiv:1611.09182 [hep-th].
\bibitem{Furey:2018yyy}
C.~Furey,
``Three generations, two unbroken gauge symmetries, and one eight-dimensional algebra,''
Phys.\ Lett.\ B {\bf 785} (2018) 84.








\bibitem{Aaij:2019wad}
R.~Aaij {\it et al.} [LHCb Collaboration],
``Search for lepton-universality violation in $B^+\to K^+\ell^+\ell^-$ decays,''
Phys.\ Rev.\ Lett.\  {\bf 122} (2019) no.19,  191801
doi:10.1103/PhysRevLett.122.191801
[arXiv:1903.09252 [hep-ex]].




\bibitem{Jungman:1995df}
G.~Jungman, M.~Kamionkowski and K.~Griest,
``Supersymmetric dark matter,''
Phys.\ Rept.\  {\bf 267} (1996) 195
doi:10.1016/0370-1573(95)00058-5
[hep-ph/9506380].

\bibitem{ArkaniHamed:2008qn}
N.~Arkani-Hamed, D.~P.~Finkbeiner, T.~R.~Slatyer and N.~Weiner,
``A Theory of Dark Matter,''
Phys.\ Rev.\ D {\bf 79} (2009) 015014
doi:10.1103/PhysRevD.79.015014
[arXiv:0810.0713 [hep-ph]].





\bibitem{Connes:1996gi}
A.~Connes,
``Gravity coupled with matter and foundation of noncommutative geometry,''
Commun. Math. Phys. \textbf{182} (1996), 155-176
doi:10.1007/BF02506388
[arXiv:hep-th/9603053 [hep-th]].
\bibitem{Chamseddine:2012sw}
A.~H.~Chamseddine and A.~Connes,
``Resilience of the Spectral Standard Model,''
JHEP \textbf{09} (2012), 104
doi:10.1007/JHEP09(2012)104
[arXiv:1208.1030 [hep-ph]].
\bibitem{Chamseddine:1996zu}
A.~H.~Chamseddine and A.~Connes,
``The Spectral action principle,''
Commun. Math. Phys. \textbf{186} (1997), 731-750
doi:10.1007/s002200050126
[arXiv:hep-th/9606001 [hep-th]].
\bibitem{Chamseddine:2014uma}
A.~H.~Chamseddine, A.~Connes and V.~Mukhanov,
``Geometry and the Quantum: Basics,''
JHEP \textbf{12} (2014), 098
doi:10.1007/JHEP12(2014)098
[arXiv:1411.0977 [hep-th]].


\end{thebibliography}
\end{document}